\documentclass{csmagclass}														
\usepackage[utf8]{inputenc}
\usepackage{amsmath}
\usepackage{amssymb}
\usepackage{subcaption} 
\DeclareUnicodeCharacter{FB01}{fi}
\begin{document}		
\headings																															 %

\title{{ Phase Diagram of a Generalized $XY$ Model with Geometrical Frustration}}
\author[1]{{ M. LACH\thanks{Corresponding author: matus.lach@student.upjs.sk}}}
\author[1]{{M. \v{Z}UKOVI\v{C}}}

\affil[1]{{ Institute of Physics, Faculty of Science, P. J. \v{S}af\'arik University, Park Angelinum 9, 041 54 Ko\v{s}ice, Slovakia}}

\maketitle

\begin{Abs}
In the present study we investigate the effects of geometrical frustration on the $XY$ model with antiferromagnetic (AFM) coupling on a triangular lattice, generalized by the inclusion of a third-order antinematic term (AN3). We demonstrate that at non-zero temperatures such a generalization leads to a phase diagram consisting of three different quasi-long-range ordered (QLRO) phases. Compared to the model with the second-order AN coupling (AN2), it includes besides the AFM and AN3 phases which appear in the limits of relatively strong AFM and AN3 interactions, respectively, an additional complex noncollinear QLRO phase at lower temperatures wedged between the AFM and AN3 phases. This new phase originates from the competition between the AFM and AN3 couplings, which is absent in the model with the AN2 coupling.
\end{Abs}
\section{Introduction}

Despite the rigorously proven absence of any true long-range ordering \cite{ref1}, the two-dimensional $XY$ model is known to exhibit an unusual infinite order phase transition belonging to the Kosterlitz-Thouless (KT) universality class \cite{ref2}. Introduction of a nematic coupling into the Hamiltonian leads to an additional phase transition between the magnetic and nematic phases, belonging in the Ising universality class \cite{ref3}. Recently, it has been shown that higher-order harmonics can lead to a qualitatively different phase diagram, with additional quasi-long-range ordered (QLRO) phases originating from the competition between the ferromagnetic (FM) and $q$-th-order (pseudo) nematic ($N_q$, $q > 2$) couplings \cite{ref4}. The new phase transitions were identified to belong to the 3-state Potts, Ising, or KT universality classes. The simplest generalization involving the second-order AN2 coupling, in addition to the AFM one, has been shown to display, on a geometrically frustrated triangular lattice, besides the AFM and AN2 phases, also an additional chiral phase above the KT line \cite{ref5}. Here we modify this model by considering the AN3 term of the third- instead of the second-order AN2 and study how the phase diagram is affected by this change. Recent investigations of the ground-state properties of such a model suggested an interesting behavior with potential interdisciplinary applications \cite{ref6}.
\section{Model and Methods}

The Hamiltonian of the generalized $XY$ model, including the $q$-th-order couplings, can be written as follows:
\begin{equation}
\mathcal{H} = J_1\sum_{\langle i,j\rangle} \cos(\phi_i - \phi_j) + J_q\sum_{\langle i,j\rangle} \cos[ q(\phi_i - \phi_j)],
\end{equation}   
where $\phi_i  \in [0,2\pi]$ represents the $i$-th site spin angle in the $XY$ plane, $J_1$ and $J_q$ are exchange interaction parameters and $\langle i,j\rangle$ denotes the sum over nearest-neighbor spins. The first term $J_1$ is a usual magnetic, i. e. FM ($J_1 < 0$)  or AFM ($J_1 > 0$) coupling, while the second term $J_q$ represents a generalized nematic, $N_q$ ($J_q < 0$) or $AN_q$ ($J_q > 0$) interaction. We consider the model (1) for $q = 3$ and the interaction parameters $J_1, J_q \in[0,1]$  in the form $J_1 = x$, $J_q = 1-x$, with $x \in \{0, 0.1, 0.2, …,1\}$ to cover the interactions between the pure AN3 $(x = 0)$ and the pure AFM $(x = 1)$ limits.

Monte Carlo (MC) simulations, based on the standard Metropolis algorithm, implemented on graphical processing units, were employed to simulate the studied system. We considered the system of a linear size $L = 96$, with periodic boundary conditions to eliminate boundary effects. The simulations were carried out for the whole relevant temperature range from $T = 0.01$, which approximates ground-state conditions, all the way to $T = 0.52$ corresponding to the paramagnetic phase. At each temperature step $10^5$ MC sweeps were used to ensure equilibration of the system and another $5\times10^5$ MC sweeps were used to calculate mean values of the following relevant quantities: the internal energy per spin
\begin{equation}
 e =\frac{\langle \mathcal{H} \rangle}{L^2},
 \end{equation}
 the specific heat per spin
\begin{equation}
C = \frac{\langle \mathcal{H}^2\rangle  -  \langle \mathcal{H}\rangle^2}{T^2 L^2},
\end{equation}
 the magnetic $(m_1)$ and generalized nematic $(m_3)$ order parameters

\begin{equation}
m_k = \frac{\langle M_k\rangle}{L^2} = \frac{1}{L^2} \left\langle  \sqrt{3 \sum_{\alpha = 1}^3 \textbf{M}^2_{k\alpha}} \right\rangle, k = 1, 3; \alpha = 1, 2, 3;
\end{equation}
where $\textbf{M}_{k\alpha}$ is the $\alpha$-th sublattice order parameter vector given by
\begin{equation}
\textbf{M}_{k\alpha} =  \left(\sum_{i \in \alpha} \cos(k\phi_{\alpha i}),\sum_{i \in \alpha} \sin(k\phi_{\alpha i})\right),
\end{equation}   
and finally, the standard ($\kappa_1$) and generalized ($\kappa_3$) staggered chiralities
\begin{equation}
\kappa_k = \frac{\langle K_k\rangle}{L^2}= \frac{1}{2L^2}\left\langle \left| \sum_{p^+\in\bigtriangleup} \kappa_{kp^+} -  \sum_{p^-\in\bigtriangledown} \kappa_{kp^-} \right| \right\rangle, k = 1, 3;
\end{equation}
where $\kappa_{kp^+}$ and $\kappa_{kp^-}$ are the local generalized chiralities for each elementary plaquette of upward and downward triangles, respectively, defined by:
\begin{equation}
\kappa_{kp} = 2\{\sin[k(\phi_2-\phi_1)] + \sin[k(\phi_3-\phi_2)]+ \sin[k(\phi_1-\phi_3)]\}/3\sqrt{3}.
\end{equation}
\section{Results}

Anomalies (peaks) in the specific heat measurements were used to determine temperatures at which the studied system undergoes phase transitions, yielding the phase diagram. The phases themselves are then characterized by order parameters, defined in the previous section. Temperature dependencies of the generalized magnetic, nematic and chiral order parameters as well as the specific heat are displayed in Fig. 1, for the values of $x = 0.2$, $0.6$, and $0.8$. 
\begin{figure}
\centering
\begin{subfigure}{.32\linewidth}
    \centering
    \includegraphics[width=\linewidth]{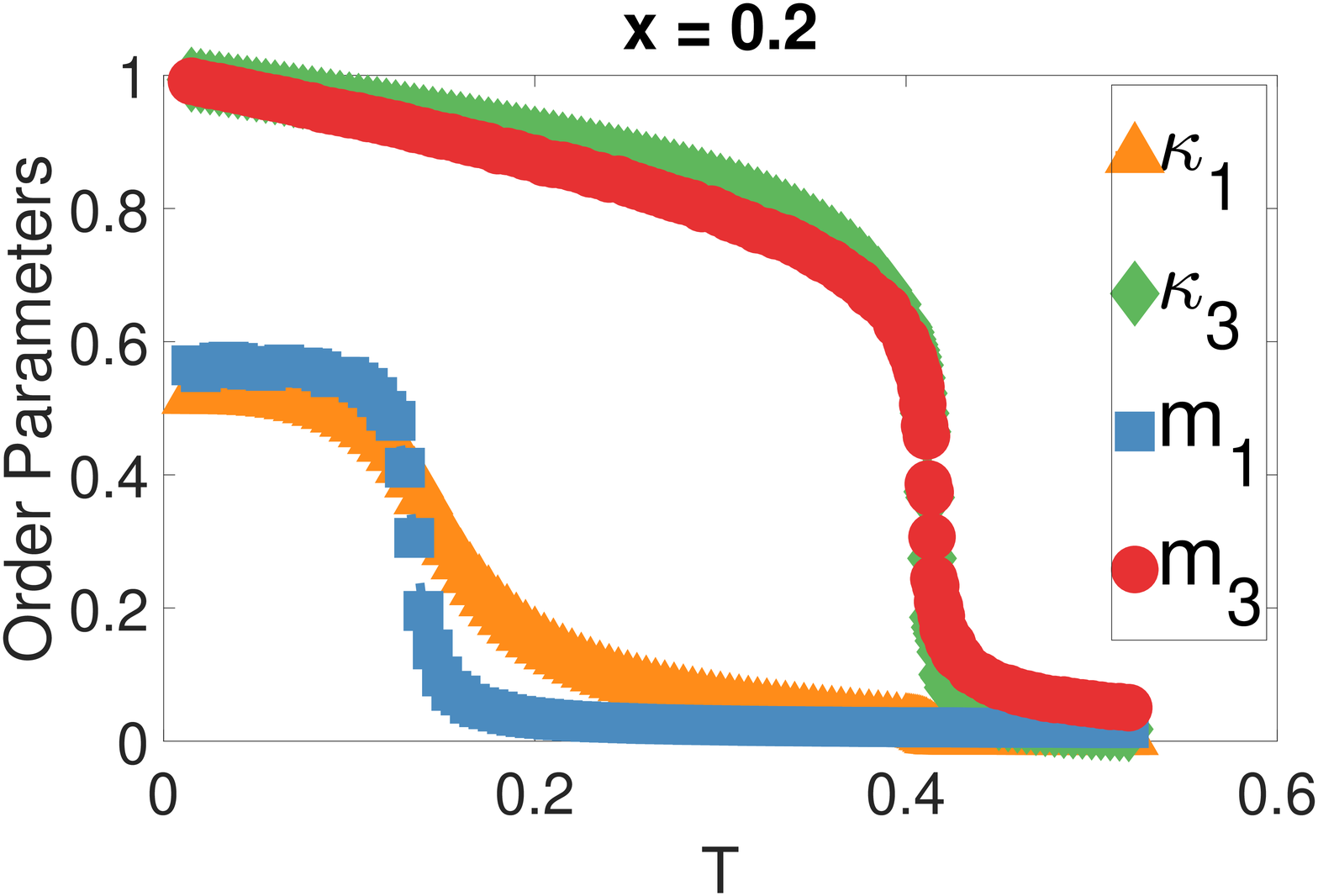}
    \caption{}\label{fig:image11}
\end{subfigure}
    \hfill
\begin{subfigure}{.32\linewidth}
    \centering
    \includegraphics[width=\linewidth]{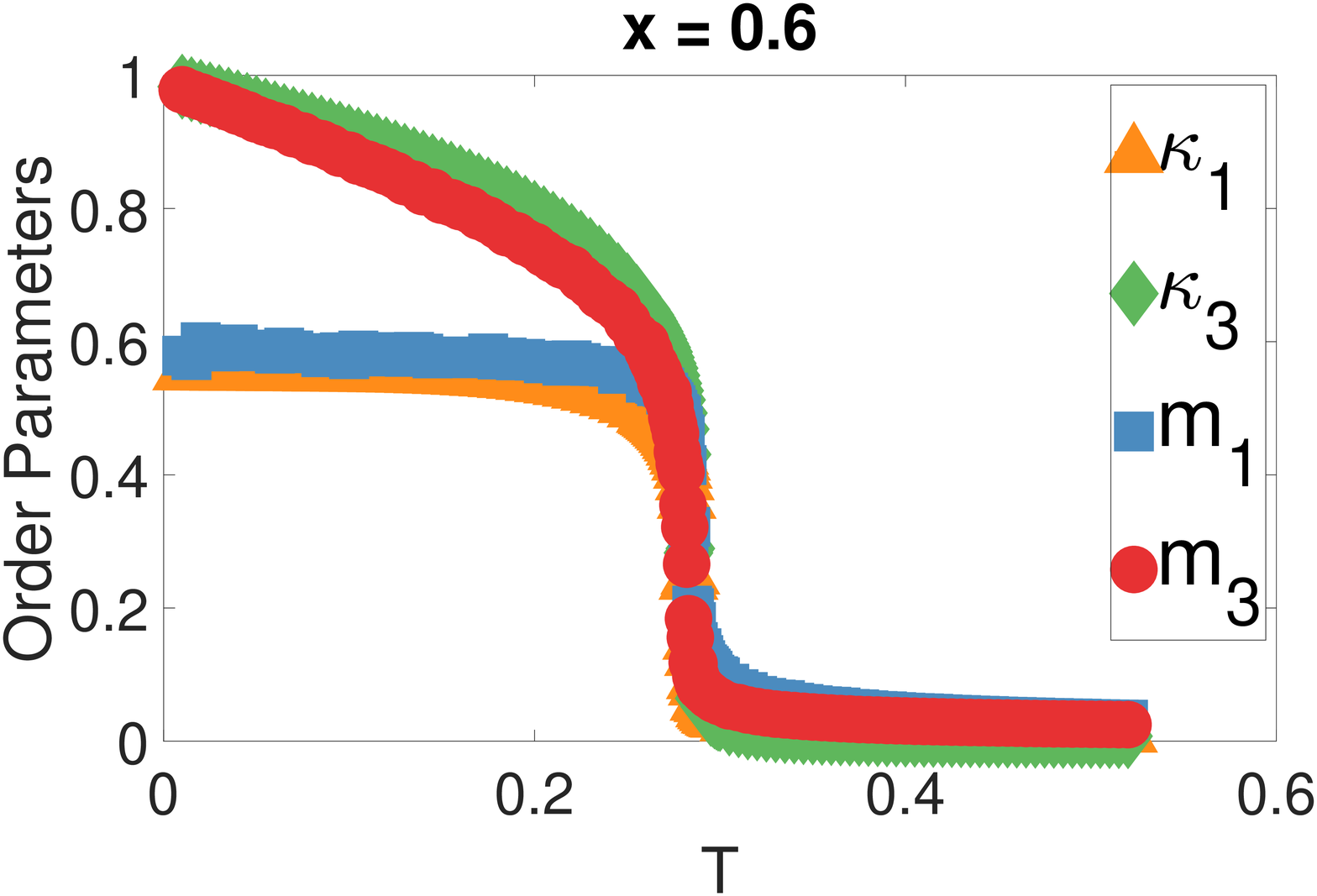}
    \caption{}\label{fig:image12}
\end{subfigure}
   \hfill
\begin{subfigure}{.32\linewidth}
    \centering
    \includegraphics[width=\linewidth]{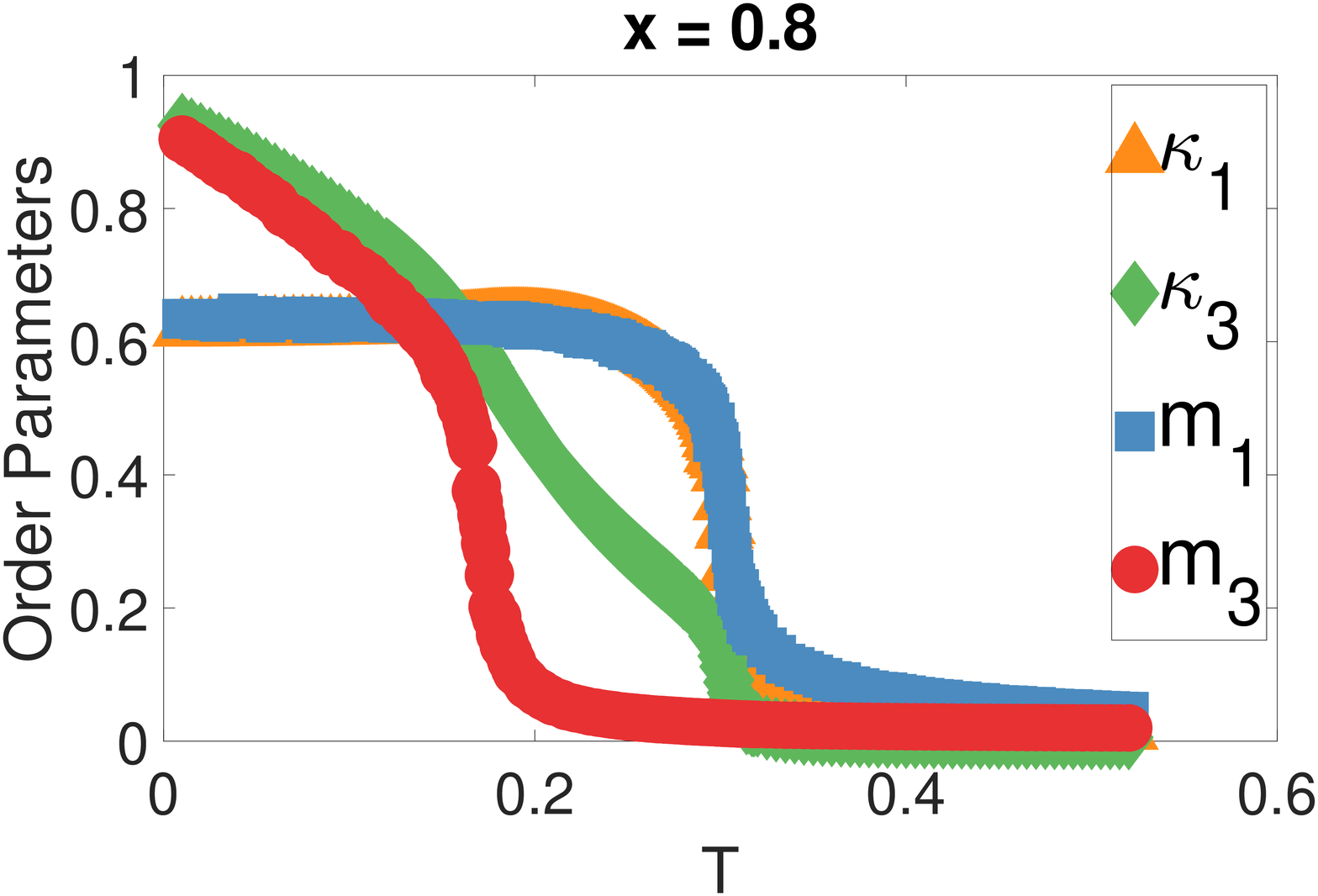}
    \caption{}\label{fig:image13}
\end{subfigure}

\bigskip
\begin{subfigure}{.32\linewidth}
    \centering
    \includegraphics[width=\linewidth]{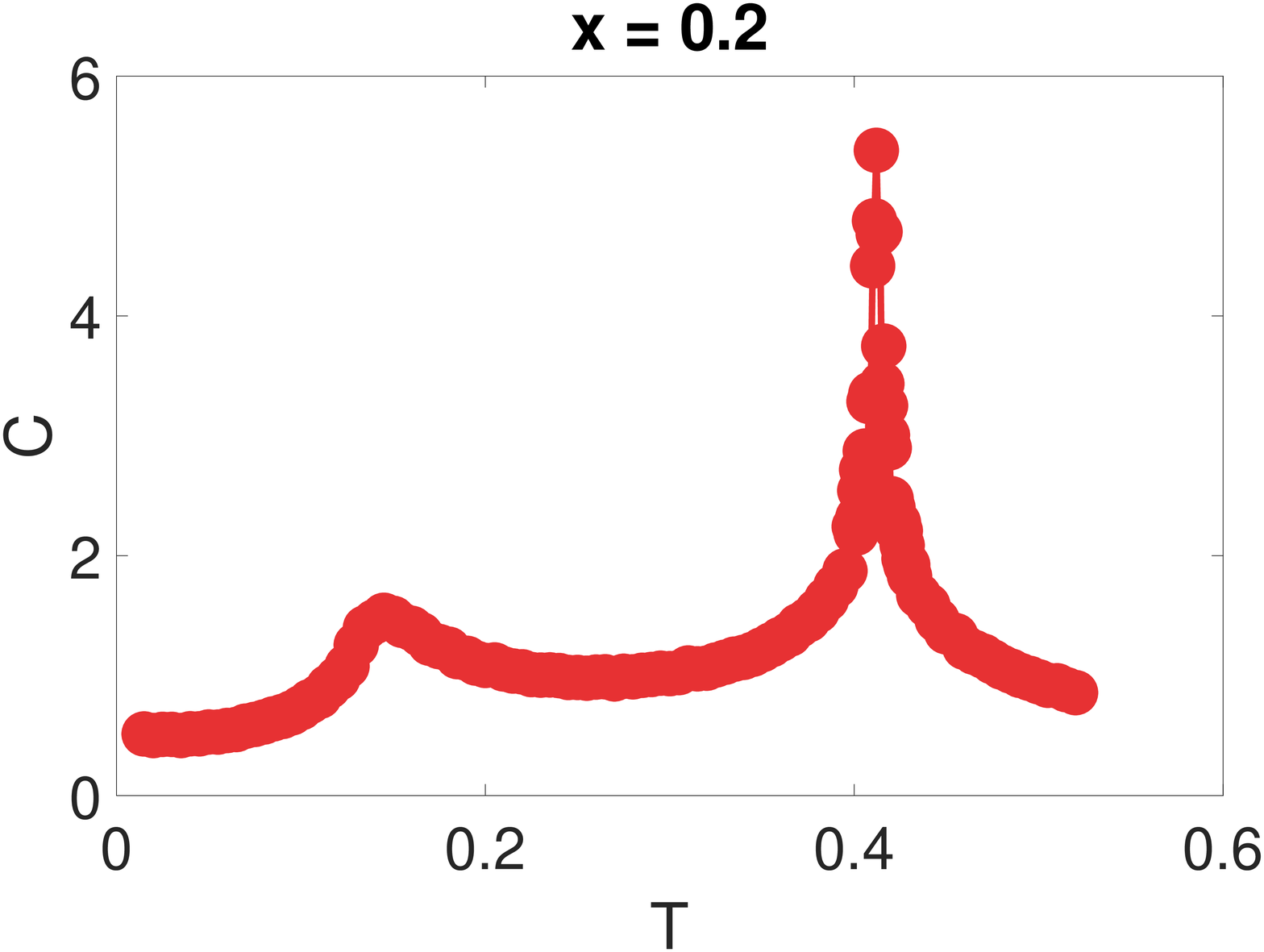}
    \caption{}\label{fig:image21}
\end{subfigure}
    \hfill
\begin{subfigure}{.32\linewidth}
    \centering
    \includegraphics[width=\linewidth]{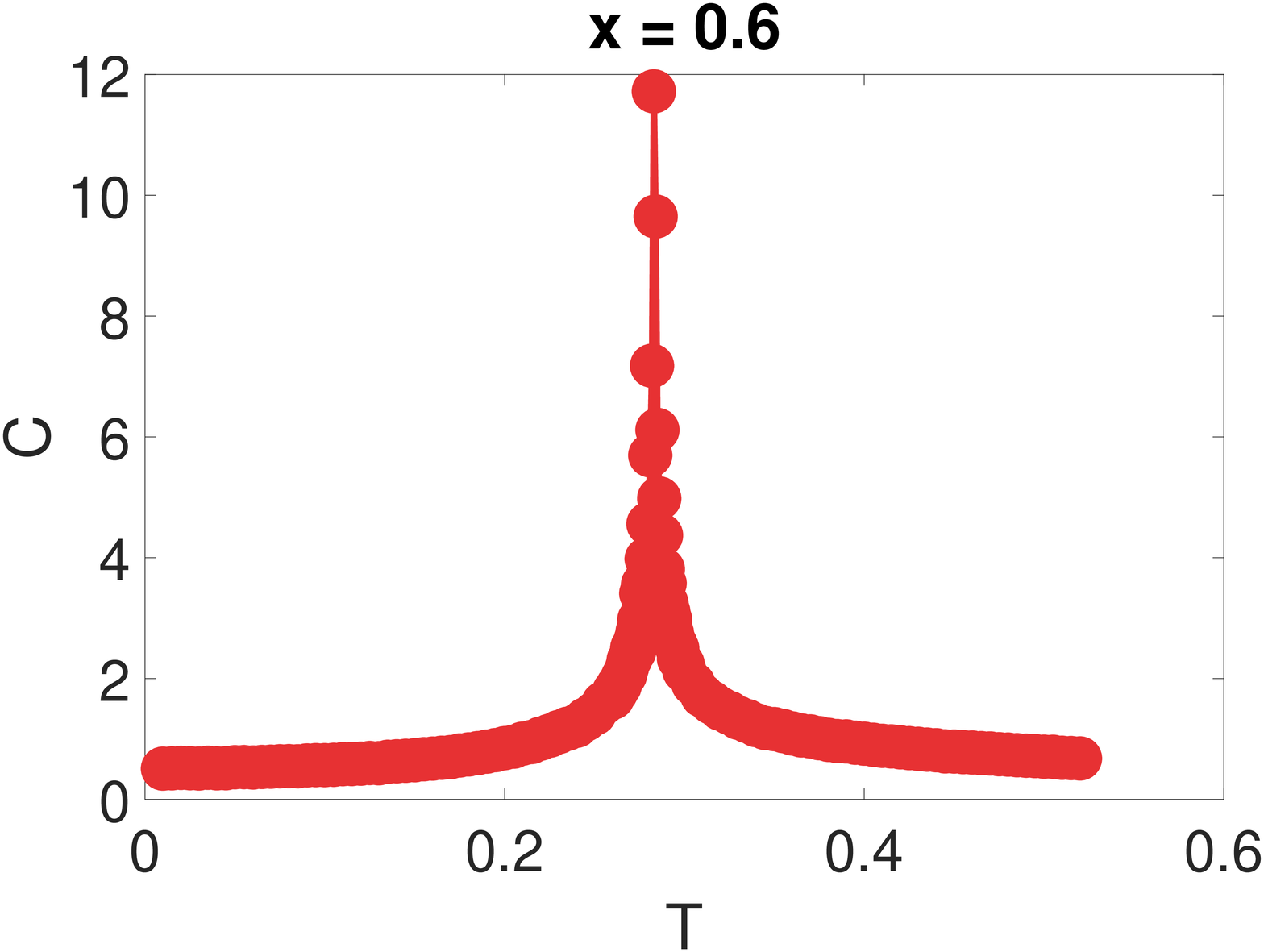}
    \caption{}\label{fig:image22}
\end{subfigure}
   \hfill
\begin{subfigure}{.32\linewidth}
    \centering
    \includegraphics[width=\linewidth]{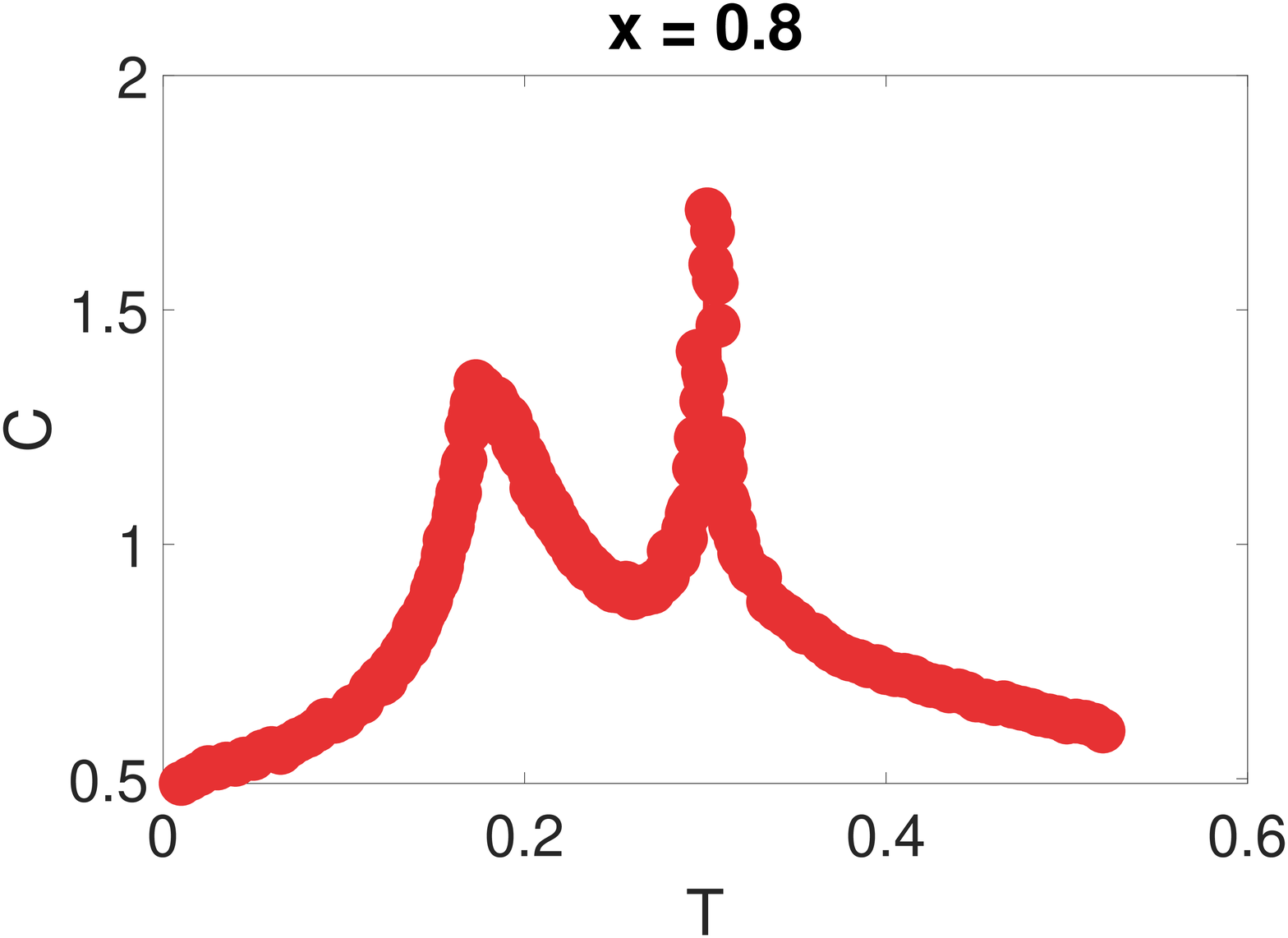}
    \caption{}\label{fig:image23}
\end{subfigure}
\caption{Temperature dependencies of different order parameters per spin (upper row) and the specific heat per spin (lower row), for three representative points in the exchange interaction parameter space.}
\label{fig1}
\end{figure}
It is clear that for $x = 0.2$ and $x = 0.8$, the magnetic $(m_1)$ and generalized nematic $(m_3)$ order parameters vanish at different temperatures. This means that for these values of the exchange interaction parameters (and as shown in Fig. 2 also in their vicinity) there are two distinct QLRO phases. At low temperatures near the ground state there is a QLRO phase in which all of the order parameters are non-zero, although, only the parameters associated with the AN3 ordering reach saturation and only for $x < 0.8$. This is due to geometrical frustration and competition between the AFM and AN3 interactions. The ground states of this model have been thoroughly investigated in Ref. \cite {ref6} and the spins on each triangular plaquette were found to be arranged in such a way that two neighbors are oriented almost parallel with respect to each other and almost anti-parallel with respect to the third one, with the turn angles dependent on the interaction strength ratio.  In the following we will refer to this phase as a canted AFM (CAFM) phase. As temperature increases to the value of the first phase transition either magnetic (for $x \lesssim 0.5$)  or nematic (for $x \gtrsim 0.6$) order parameter falls to zero while the corresponding chiral order parameter shows an anomalous decrease, but remains non-zero. In the second QLRO phase this chiral order parameter continues to decline, but stays slightly above zero all the way until the second phase transition to the paramagnetic state. The other two parameters - nematic for $x \lesssim 0.5$ and magnetic for $x \gtrsim 0.6$ and their corresponding chiral order parameters decrease slightly but remain largely unaffected until the transition to the paramagnetic state where all the order parameters vanish. The presence of three distinct phases is further supported by our calculations of the specific heat per spin (Fig. 1 lower row), which clearly displays two peaks at two separate temperatures corresponding to the drops of order parameters, as described above.

For $0.5\leq x\leq 0.6$ the situation changes in the way that the CAFM phase persists as the temperature is increased until the system undergoes a transition directly to the paramagnetic state with all the order parameters vanishing together. In this case there is only a single peak in the specific heat, corresponding to this transition.
\begin{figure}[h!]
\centering
\includegraphics[width=8cm]{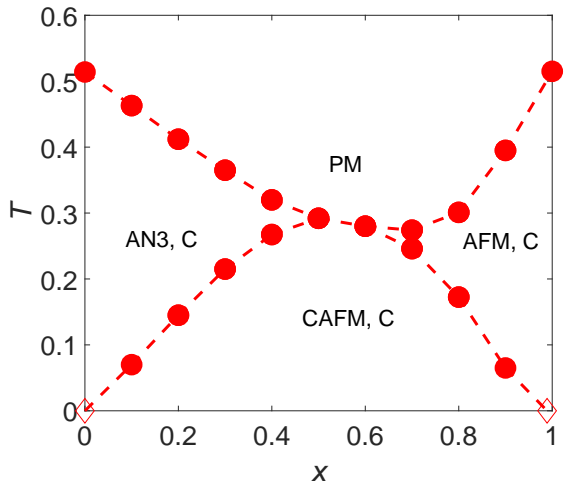}
\caption{Phase diagram in the $x-T$ parameter plane. The symbols represent temperatures corresponding to the maxima of the specific heat, lines serve only as a guide to the eye. Empty symbols represent the limits of the CAFM phase obtained from the ground-state analysis conducted in \cite{ref6}.}
\label{fig2}
\end{figure}
The phase diagram depicted in Fig. 2 covers the whole range of the exchange parameter space from the purely AFM ($x = 1$) to the purely AN3 ($x = 0$) cases. The behavior in the limiting cases is well known - there is a single phase transition from the AFM, for $x = 1$ or AN3, for $x = 0$, phases, respectively, to the disordered paramagnetic state at higher temperatures. For $0.0 < x \lesssim 0.997$ (see Ref. \cite{ref6}) there is a CAFM phase at low temperatures which gives way to  the AN3 phase ($0.0\lesssim x \lesssim 0.5$), AFM phase ($0.6 \lesssim x \lesssim 0.997$) or straight to the paramagnetic phase ($0.5\lesssim x \lesssim 0.6$). It should be noted, that the transition to the paramagnetic phase occurs at much lower temperatures compared to the purely AFM and AN3 cases.
\section{Conclusions}

We have studied the effects of geometrical frustration and competition between the AFM and AN3 couplings in a generalized $XY$ model. In the work of Poderoso et. al. \cite{ref4}, which studied the corresponding non-frustrated model with the ferromagnetic and nematic interactions, the inclusion of $q-th$-order nematic couplings leads to new ordered phases for $q \geq 5$. In contrast, in the present model we observe the emergence of a new CAFM phase already for $q = 3$. This phase, not present in the case of $q = 2$ \cite{ref5}, is characterized by chiral, AFM and AN3 ordering with only the parameters corresponding to AN3 interaction for $x < 0.8$ reaching saturation. For roughly equal strength of the AFM and AN3 interactions, the competition forces the system to transition directly from the CAFM into the paramagnetic state at relatively low temperatures. The transitions to the paramagnetic phase are believed to belong to the KT universality class \cite{ref5}, whereas the nature of the transitions between the CAFM phase and AFM / AN3 ordered phases is not yet precisely known. The reason is a high degree of frustration and competition, which makes it difficult to obtain statistically significant results from standard MC simulations at critical temperatures. Further study using more sophisticated methods is desirable.
\section{Acknowledgement}
This work was supported by the Scientific Grant Agency of Ministry of Education of Slovak Republic (Grant No. 1/0531/19).
%


\end{document}